body of paper
\def\be{\begin{equation}}
\def\te{\end{equation}}
\def\bea{\begin{eqnarray}}

\def\tea{\end{eqnarray}}

\def\d{\delta}

\def\S{\Sigma}

\newskip\humongous \humongous=0pt plus 1000pt minus 1000pt

\newif\ifdtup

\documentstyle[12pt]{article}

\makeatletter                    
\@addtoreset{equation}{section}  
\makeatother                     

\begin{document}
\title{Environment-Induced Effects on Quantum Chaos: \\
 Decoherence, Delocalization and Irreversibility
\thanks{
Invited parallel session talk by K. Shiokawa at the International Symposium
on Quantum Classical Correspondence, Drexel University, Philadelphia,
Sept. 8-11, 1994.
To appear in the Proceedings edited by D. H. Feng and B. L. Hu
(International Publishers, Boston, 1995)}}
\author{K. Shiokawa $^1$ and B. L. Hu $^{1,2}$ \\
{\small $^1$ Department of Physics, University of Maryland, College Park,
MD 20742, USA}\\
{\small $^2$ School of Natural Sciences, Institute for Advanced Study,
Princeton,
NJ 08540, USA}}
\date{\small {\it (IASSNS-HEP-95/3, UMDPP 95-081, January 7, 1995)}}
\maketitle

\begin{abstract}
Decoherence in quantum systems which are classically chaotic is studied.
It is well-known that a classically chaotic system when quantized
loses many prominent chaotic traits. We show that interaction of the
quantum system with an environment can under general circumstances quickly
diminish quantum coherence and reenact some characteristic classical
chaotic behavior. We use the Feynman-Vernon influence functional
formalism to study the effect of an ohmic environment at high temperature
on two classically-chaotic systems: The linear Arnold cat map (QCM)
and the nonlinear quantum kicked rotor (QKR).
Features of quantum chaos such as recurrence in QCM and diffusion suppression
leading to localization in QKR are destroyed in a short time due to
environment-induced decoherence.
Decoherence also undermines localization and induces
an apparent transition from reversible to irreversible dynamics
in quantum chaotic systems.

\end{abstract}


\section{Decoherence and Disappearance of Recurrence in the Quantum Cat Map}

Arnold's cat map is a linear area-preserving map $T$ on a torus in phase
space formed by
identifying the boundaries of the interval $[0,2 \pi]$
in both the coordinate $Q$ and the momentum $P$ directions \cite{ArnAve}.
(Because of this 
the area of the torus is characterized by Planck's constant which
takes on the  values $\hbar = 2 \pi / {\cal N} $,
where ${\cal N}$ is the number of sites in both the coordinate and the momentum
directions in the phase space.)
 From time step $j$ to $j+1$ it is given by
\begin{equation}
    \left( \begin{array}{c}
		Q_{j+1}\\
		P_{j+1}
	   \end{array}      \right)  =
    \left( \begin{array}{cc}
		a & b \\
		c & d
	   \end{array}      \right)
    \left( \begin{array}{c}
		Q_j\\
		P_j
	   \end{array}      \right)  =  T
     \left( \begin{array}{c}
		Q_j\\
		P_j
	   \end{array}      \right)
\end{equation}
where $det T = 1$ guarantees area preservation.
The degree of chaos depends on the choice of $T$.
The eigenvalues of $T$ are either both real or both imaginary.
In the latter case, $T$ is elliptic, the motion becomes periodic
and no sensitive dependence on the initial
condition is observed. When $T$ is hyperbolic, the motion is chaotic.

Quantized cat map is studied in detail by Hannay and Berry \cite{HanBer}.
The matrix has to assume a special form in order to yield nontrivial values
of the progagator for the map. We choose
\begin{equation}
     T_{1} = \left( \begin{array}{cc}
		 0 & -1 \\
		 1 &  0
	   \end{array}      \right),~~
     T_{2} = \left( \begin{array}{cc}
		 2 & 3\\
		 1 & 2
	   \end{array}      \right)
\end{equation}
for the elliptic and the hyperbolic cases respectively.

For the special choice of the matrix elements $T_1, T_2$ made above
the propagator takes on the simple forms,
\begin{equation}
U_1(j+1, j) = \sqrt{\frac{i}{\cal N}}\exp[-\frac{i}{\hbar}Q_{j}Q_{j+1}], ~~
U_2(j+1, j) = \sqrt{\frac{i}{\cal N}}\exp[ \frac{i}{\hbar}
(Q_{j}^{2}- Q_{j}Q_{j+1}+Q_{j+1}^{2})].
\end{equation}
Since each iteration describes a permutation among sites,
each site belongs to a periodic orbit. Thus the quantum dynamics
follows the classical way, resulting in the recurrence of the wave function
(or equivalently, the Wigner function \cite{HanBer}).

We now couple the system linearly to a bath of $N$ harmonic oscillators
with coordinates $q_\alpha$  and momentum $p_\alpha$ ($ \alpha = 1,..N $)
described by the Hamiltonian $H_B$ and the interaction Hamiltonian $H_C$
\begin{equation}
 H_{B} = \sum_{\alpha=1}^{N}( \frac{p_{\alpha}^{2}}{2}
	      + \frac{\omega_{\alpha}^{2}q_{\alpha}^{2}}{2}), ~~
 H_{C} = \sum_{\alpha=1}^{N} C_{\alpha} Q q_{\alpha}.
\end{equation}
where $Q$ is the coordinate of the system and $C_\alpha$ is the coupling
constant of $Q$ to the $\alpha$th oscillator in the bath.
By integrating out the bath variables, we get the reduced density matrix,
\begin{equation}
  \rho_r(Q_{j}, Q'_{j}, t) =
      \int \Pi_{\alpha=1}^{N} dq_{\alpha} dq'_{\alpha} \exp \frac{i}{\hbar}
      [S(Q)+S_C(Q,q_{\alpha})+S_B(q_{\alpha})
      -S(Q')-S_C(Q',q'_{\alpha})- S_B(q'_{\alpha})].
\end{equation}
where $S$ is the classical action of the system which appears as the exponent
 of the propagator in (1.3).
$S_B$, and $S_C$ are the actions for bath and
interaction,  
respectively.
The propagator $J_r$ for the reduced density matrix
from time steps $j$ to $j+1$ is
\begin{equation}
  J_r(Q_{j+1}, Q'_{j+1} \mid Q_{j}, Q'_{j}, t) =
      \int DQ DQ' \exp \frac{i}{\hbar}[S(Q)-S(Q')+ A(Q, Q')],
\end{equation}
in a path-integral representation \cite{FeyVer,CalLeg83,HPZ1}, where
\begin{equation}
\frac{i}{\hbar}A(Q, Q')= \frac{1}{\hbar^{2}}\int_{0}^{t} ds \int_{0}^{s} ds'
 r(s)[-i \mu(s-s')R(s') - \nu(s-s') r(s')]
\end{equation}
is the influence action.
Here  $ r \equiv \frac{Q - Q'}{2},
 R \equiv \frac{Q + Q'}{2}$, and $\mu(s), \nu(s)$ are
the dissipation and noise kernels respectively \cite{HPZ1}.

If we consider the simplest case  of an ohmic bath at high temperature
 $kT > \hbar \Lambda >> \hbar \omega_\alpha$ \cite{CalLeg83},
 and consider times shorter than the relaxation time,
then we obtain a Gaussian form for the influence functional, with
$    {\frac{i}{\hbar}A(Q, Q')} =
     {-\frac{2 M \gamma k T}{\hbar^{2}} \Sigma_{j} r_{j}^{2}}.
$
where the noise kernel becomes local $\nu(s) = 2 M \gamma k T \delta(s)$
and $\gamma$ is the damping coefficient.
The unit-time propagator becomes
\begin{equation}
  J_{r}(Q_{j+1}, Q'_{j+1} \mid Q_{j}, Q'_{j}) = \langle
  J_{r}(Q_{j+1}, Q'_{j+1} \mid Q_{j}, Q'_{j}, \xi)  \rangle
 =\langle \exp \frac{i}{\hbar} [S(Q_{j+1},Q_{j})- S(Q'_{j+1},Q'_{j})
				+\xi r_{j+1}] \rangle.
\end{equation}
Here $\xi$ is a Gaussian white noise given by
\begin{equation}
	    \langle \xi \rangle = 0 , ~~~
	    \langle \exp \frac{i}{\hbar} \xi r \rangle
	    = \exp [   -\frac {2 M \gamma k T}{\hbar^{2}} r^2 ]
\end{equation}
where $\langle ~\rangle$ denotes statistical average over noise realization
$\xi$.

For the elliptic map, we get
\begin{equation}
  J_{r}(Q_{j+1}, Q'_{j+1} \mid Q_{j}, Q'_{j}, \xi)
 = (\frac{i}{\cal N})^{1/2}
 \exp [\frac{i}{\hbar}  (- r_{j} R_{j+1} - r_{j+1} R_{j} + \xi r_{j+1})].
\end{equation}
and for the hyperbolic map,
\begin{equation}
  J_{r}(Q_{j+1}, Q'_{j+1} \mid Q_{j}, Q'_{j}, \xi)
 = (\frac{i}{\cal N})^{1/2}
 \exp [ \frac{i}{\hbar}
	(2 r_{j} R_{j} + 2 r_{j+1} R_{j+1}
	- r_{j} R_{j+1} - r_{j+1} R_{j} + \xi r_{j+1} )]
\end{equation}

The Wigner function is defined as
\begin{equation}
  W(R,p) = \frac{1}{\pi \hbar}
	  \int_{0}^{2 \pi} \psi (R+r) \psi^{*} (R-r)
	   \exp ( \frac {2i}{\hbar} pr) dr.
\end{equation}
where $p$ is the momentum conjugate to $r$.
The propagator $K$ for the Wigner function is
\begin{equation}
    \begin{array}{l}
      K(R_{j+1}, p_{j+1}\mid R_{j}, p_{j}, \xi)
     =  \Sigma_{r_{j}} \Sigma_{r_{j+1}}
     J_{r}(Q_{j+1}, Q'_{j+1} \mid Q_{j}, Q'_{j}, \xi)
      \exp  \frac{2 i}{\hbar} ( p_{j} r_{j} - p_{j+1} r_{j+1}).
    \end{array}
\end{equation}
This is reduced to the form of the classical cat map.
For the elliptic case,
 \begin{equation}
R_{j} = -p_{j+1} + \xi, ~~~ p_{j} =  R_{j+1} .
\end{equation}
For the hyperbolic case,
\begin{equation}
R_{j} = -p_{j+1} + 2R_{j+1} + \xi , ~~~ p_{j} = - 3R_{j+1} + 2p_{j} -2\xi.
\end{equation}

Without noise, quantum evolution follows classical permutation \cite{HanBer}
the phase space is divided by a finite number of different
periodic orbits and the period is known to increase roughly
proportional to ${\cal N}$ with some irregular oscillation.
When coupled to a bath, the cat map  is exposed to a Gaussian noise in each
time step. The discretized noise induces transitions between different
periodic orbits in an irregular way.
Interaction with an environment blurs the recurrence of
physical quantities in the quantum map.
Fig.1 shows $Tr \rho_r^{2}$, the linearized entropy (with the
reversed sign) for various cases.
If there is no interaction with the environment,
the entropy is constant for both regular and chaotic cases.
Quantum recurrence is evident even when the system is
chaotic.
When interaction sets in, $Tr \rho_r^{2}$ decays
exponentially, showing that the system rapidly decoheres.
The rate of decoherence is much faster in chaotic systems than in
 regular systems \cite{TamSip}.
It suggests that recurrence would be less evident in a decohering chaotic
system. In Fig. 2, we show the mean displacement of points
in the phase space as a function  of time steps.
This is defined by
$l = \sqrt{\langle \Delta x^2 + \Delta p^2 \rangle}$, where $\Delta x$ and
$\Delta p$ are the displacements from the initial phase space points,
and  $\langle~\rangle$ denotes averaging over noise distributions.
In the chaotic case, we see that recurrence disappears with just a small
amount of noise (Fig. 2a) whereas in the regular case,
the same amount of noise does not alter
the qualitative picture of recurrence (Fig. 2b).
In both cases, the decohered quantum system behaves close to the classical
picture in which
the regular and chaotic dynamics are clearly distinguished.
In spite of the discreteness of the points on the torus,
the system  behaves effectively classically due to the influence of
the environment.

\section{Decoherence and Delocalization in the Quantum Kicked Rotor}
\setcounter{equation}{0}

The kicked rotor is one of the most
intensively studied models from both the quantum and classical point of view
\cite{LicLib}.
 The Hamiltonian of the kicked rotor is given by
\begin{equation}
  H = \frac{p^{2}}{2m}  + K \cos x \Sigma_{j=-\infty}^{\infty} \delta(t - j)
\end{equation}
which describes  a one-dimensional rotor subjected to a delta-functional
periodic kick at $t=j$.
Here $x$ is the angle of the rotor with period $2 \pi$,
$m$ is the moment of inertia, $p$ is the angular momentum,
and  $K$ is the strength of the kick which measures the nonlinearity.
When $K > 1$, the system becomes chaotic over the entire phase space.

The quantum dynamics of the kicked rotor is given by the corresponding
Schr\"{o}dinger equation
\begin{equation}
     i \hbar \frac{\partial}{\partial t}
 \psi (x,t) = -\frac{\hbar^{2}}{2 m}\frac{\partial^{2}}{\partial x^{2}}
 \psi (x,t) + K \cos x \S_j \d (t-j) \psi (x,t)
\end{equation}
where $\psi$ is the wave function of the rotor.

Denoting $\psi_{j}$ as the wave function $\psi(x,t)$ at each discrete
time $t = j$,
and integrating (2.2) from $j$ to $j+1$, we obtain
\begin{equation}
  \psi_{j+1} (x) =
 \exp [-i \frac{\hbar}{2m}  \frac{\partial^2}{\partial x^2}]
 \exp [-i \frac{K\cos x}{\hbar} ]
 \psi_{j}(x)
\end{equation}

The quantum kicked rotor (QKR) is known to exhibit dynamical localization.
After some relaxation time scale,
the wave function becomes exponentially localized in the momentum space
 \cite{CCIF}.
This may be interpreted as a particle moving in a lattice with a quasi-random
potential.
This heuristic picture seems to justify the analogy between
the quantum kicked rotor to the tight binding model
with an exponentially decaying hopping parameter which is known to show
 Anderson localization \cite{FGP}. Dynamical localization in this
context arises from 
the suppression of classical diffusive behavior by the quantum dynamics.
However, as shown by Ott, et.al. \cite{OAH}, a small external noise
can break the localization.
Sufficient amount of noise would induce the quantum system to
exhibit classical diffusive behavior.
Dittrich and Graham studied this problem \cite{DitGra} by coupling
the system to a zero temperature harmonic oscillator bath
and analysed solutions to the master equation.
Cohen and Fishman presented the most detailed study of this problem
for an ohmic bath \cite{CohFis}.
Here we want to approach these issues from
an environment-induced decoherence point of view \cite{dec}.
We begin by calculating the density matrix for the kicked rotor
coupled to an environment.

 We introduce a linear coupling of the system momentum $p$
 with each oscillator coordinate
 $q_ \alpha (\alpha = 1,..N)$ in the bath in the form
 $ H_C = \Sigma_{\alpha=1}^N C_\alpha q_{\alpha} p $
(Here $q, p$ without the subscript $\alpha$ denote the system coordinate and
 momentum variables).
As before, we assume an ohmic bath and examine the time period where
dissipation is small. Under these assumptions,
 the unit time propagator for the wave function
$ U_\xi(j+1,~j) $ is given by
\begin{equation}
 U_\xi(j+1,~j)
=   \exp [ -\frac{i}{\hbar}  \frac{p^2}{2m}]
    \exp [ -\frac{i}{\hbar}  K\cos x ]
    \exp [ -\frac{i}{\hbar}  \xi p ]
\end{equation}
where, as before, the noise term $\xi $
arises from using a Gaussian
identity in the integral transform of the term involving the noise kernel
in the influence functional.
Summing over all noise realizations $\langle ~\rangle $
gives the desired reduced density matrix,
\begin{equation}
\rho_{rj}(p,~p')
=\langle {\psi_{j,}}_\xi(p) ~{\psi_{j,}}_\xi(p')\rangle_{\xi}
\end{equation}
where
\begin{equation}
{\psi_{j+1,}}_\xi(p) = U_\xi(j+1,~j) ~{\psi_{j,}}_\xi(p)
\end{equation}

Loss of quantum coherence is measured by the density matrix
becoming approximately diagonal.
$Tr \rho_r^{2}$ can be expressed as
\begin{equation}
Tr \rho_r^{2} =
 \langle
	 \Sigma_{p} \Sigma_{p'} \psi_\xi(p) ~\psi^{*}_\xi(p')
	  ~\psi_{\xi'}(p') ~\psi^{*}_{\xi'}(p)
  \rangle_{\xi,\xi'}
\end{equation}
where $ \langle ~\rangle_{\xi,\xi'} $ denotes the statistical average of
all possible noise histories
of two independent noises $ \xi,\xi'$ defined at each time interval from
$j$ to $j+1$.
At high temperatures
$\xi(\tau),\xi'(\tau)$ are reduced to two time-uncorrelated
independent Gaussian white noises
defined at each time step.

We see that there is a close relation between the breaking of
 dynamical localization and quantum decoherence.
In Fig.3 we plot the linearized entropy $Tr \rho_r^{2}$ versus
the energy $\langle p^2 \rangle$.
This shows that delocalization occurs as quantum coherence breaks down,
suggesting that delocalization and decoherence occurs by the same mechanism.
As the nonlinearity parameter $K$ increases, the system decoheres more rapidly.
At the same time, the amount of delocalization measured by the
diffusion constant increases.

This may be explained in the following way:
Because the coupling is through the momentum,
the noise term does not involve any nonlinearity.
The time scale for the system to lose coherence is given by
$ t_{D} = (\lambda_{tdB} /\delta p)^2 /\gamma $,
where $\lambda_{tdB} = h / \sqrt{2\pi m k T}$
is the thermal de Broglie wavelength, and
$\delta p$ is the relevant momentum scale.
After this time, noise will destroy the quantum coherence
between such  momentum separations.
In the kicked rotor case, localization will occur due to the
coherence around $\delta p \sim  \Delta$, where $ \Delta \sim l\hbar $
 is the localization length.
Since $l \sim K^2$, this gives $ t_{D} \sim K^{-4} $.
This shows that nonlinearity increases the rate of decoherence.

The relation between the diffusion constant $D$ and the noise strength
is given in \cite{OAH,CohFis}.
For our case, $K/\hbar \gg 1$ and for weak noises,
we can consider the particle as undergoing a random walk with
hopping parameter $1/t_c$.
Then $D = \Delta^2 / {t_D} = (\Delta^4 / \lambda_{tdB}^2)/ \gamma $.

\section{Decoherence and Irreversibility in Quantum Chaos}

The Wigner function is often used to examine the quantum to classical
transition.
\footnote{For a linear system the Wigner function is known to show a smooth
convergence
 to the classical Liouville distribution.
But if the system Hamiltonian has a nonlinear term,
quantum corrections associated with the higher derivatives of the
potential pick up the rapid oscillations in the Wigner function and it
no longer has a smooth classical limit \cite{BerHel}.
However, upon interaction with an environment,
a coarse-grained Wigner function can have have a smooth
classical limit \cite{TakHab} for nonlinear systems.}
The Wigner function at time $t=j$ is defined as
\begin{equation}
W_j(X,p)
= \frac{1}{4\pi\hbar}\int\limits_{-2\pi}^{+2\pi} dy ~ e^{{i\over \hbar} py} ~
  \rho_j(x+{y\over 2}, x-{y\over 2}),
\end{equation}
where $X \equiv \frac{1}{2}(x+x'), y \equiv x-x'$.
 From (2.3), the unit-time propagator for the Wigner function of
the QKR is found to be
\begin{equation}
W_{j+1}(X,p) =
 e^{-\frac{K\sin x}{\hbar} \Delta_p} e^{-p {\partial_x}}W_{j}(X,p)
\end{equation}
where $\Delta_p \equiv e^{\frac{h}{2} \partial_p}
- e^{-\frac{h}{2} \partial_p} $ measures the effect of the kick. We can see the
effects of quantum corrections is seen more clearly if we expand
$\Delta_p$ in orders of $\hbar$:
\begin{equation}
e^{-\frac{iK\sin x}{\hbar} \Delta_p}
\approx e^{-K\sin x  \partial_p}e^{\frac{\hbar^2}{24} K\sin x \partial_p^3}...
\end{equation}
The first exponential contains the classical propagator and the second
contains quantum corrections of even orders of $\hbar$.
Thus we get
\begin{equation}
W_{j+1}(X, p)\approx e^{\frac{\hbar^3}{24} K\sin x \partial_p^3  }
	      W_j(X - (p + K sinx), p - K sinx)
\end{equation}
where the Wigner function with the new arguments depicts classical evolution.
This map alone is the source of streching and folding of volume in
phase space which signify classical chaos.

If the initial system wavefunction is described by a Gaussian wave packet
with width $ \delta p (>> \hbar) $,
we would expect to see a
classical-like evolution of the packet at short times.
When the width of the contracting wave packet gets so small as
comparable to $\hbar$, the  effect of quantum corrections
from higher $\hbar$ order terms in (3.4) set in.
By comparing the classical and quantum terms, we see that quantum corrections
will become important when  $\delta p(t) \sim \hbar$.
Here $\delta p(t) = \delta p(0) e^{-\lambda t}$, where
the Lyapunov exponent $\lambda \sim ln (K/2)$ .
Thus we can deduce the Ehrenfest time
\footnote{Ehrenfest time is customarily defined as the time when quantum
dyanmics
can be adequately described by the classical equations, i.e., the time $t <
t_E$
when the Wigner function or the expectation value of
any observable follow classical trajectories.}
for QKR to be
$t_E \sim  \frac {1}{\lambda} ln \frac{\delta p(0)}{\hbar}$
Note that in the continuum case, this definition gives us a
different time scale for each term in the expansion \cite{ZP}.

The major effect of the bath (at times short compared with the relaxation time)
is the appearance of a diffusion term in (3.4),
\begin{equation}
W_{j+1}(X,p) \approx
  e^{  D \partial_p^2 }
  e^{    \frac{\hbar^3}{24} K\sin x \partial_p^3  }
	      W_j(X - (p - K sinx), p - K sinx)
\end{equation}
Competition amongst the three terms with different physical
origins is apparent:
The first term in (3.5) is the quantum diffusion term,
the second is the quantum correction term, and the third is purely
classical evolution.
As discussed by Zurek and Paz \cite{ZP}, if D is sufficiently large,
the effect of quantum corrections becomes inconspicuous.
In this case, the diffusion term traces out a small scale oscillating behavior
before quantum corrections have a chance to
 change classical evolution.
Then one may expect the time evolution of the Wigner
 function to be like that of classical evolution with noise.
The role of quantum diffusion is to add some Gaussian averaging
so that the contracting direction in phase space will be suppressed while
it does not affect the stretching direction.
As long as the width of the wave packet is large such that
the first term is negligible, the evolution should be Liouvillian
(time reversible if we assume infinite measurement precision).
Furthermore, we expect that after
the width of the packet along the contracting direction becomes comparable
to the diffusion generated width (in the Gaussian wave packet)
, the dynamics will start showing irreversible
behavior arising from coarse graining (as distict from irrreversibility from
instability).
Consequently, entropy should increase in this regime.
In Fig. 4a, we plot the von Neumann entropy for the dynamics of (3.5).
We can see three qualitatively different regimes:
I. the Liouville regime: the entropy is constant and the dynamics is
    time reversible.
II. the decohering regime: the entropy keeps increasing due to
    coarse graining.
III. the finite size regime: due to the bounded nature of the phase space,
    the entropy shows saturation.
Our result from quantitative analysis seems to confirm the qualitative
description of Zurek and Paz \cite{ZP} who used the inverted harmonic
oscillator
potential as a generic source of instability.
Since the phase space in their model is not bounded
they do  not see Regime III.
Similar features appear in the quantum cap map (Fig. 4b)
In this case, the full quantum dynamics can be calculated in a simple way.
Resemblance with the result of a classical rotor with noise is obvious.
However, in this case, the stable entropy
is smaller than the maximum value which may be explained as
a finite (phase space) size effect.\\

\noindent {\bf Acknowledgement}
We thank Drs. Shmul Fishman and Juan Pablo Paz for explaining their work
and Drs. Ed Ott and Richard Prange for general discussions.
Research is supported in part by the National Science Foundation
under grant PHY91-19726. BLH gratefully acknowledges support from
the General Research Board of the Graduate School of the University of
Maryland and the Dyson Visiting Professor Fund at the  Institute
for Advanced Study, Princeton. \\

\noindent{\bf Figure Captions}\\

\noindent {\bf Figure 1} The linearized entropy (with reversed sign)
$Tr \rho_r^2$ is plotted here as a function of time.
If there is no environment, the entropy is constant for both hyperbolic and
elliptic cases, indicating the purity of the state.
For the hyperbolic map,
even though classically this system is strongly chaotic,
the corresponding quantum system does not show chaotic behavior.
This situation changes drastically when the system interacts with
a thermal bath: entropy keeps increasing due to coarse graining.
Note that in the  hyperbolic case (solid line)
the rate of entropy increase is greater than in the elliptic case
(dotted line). {\cal N = 50 } is used here (also in Fig.2).\\
\noindent {\bf Figure 2} The mean phase space point displacement is shown.
When there is no environment (dotted line), the system shows  recurrence
in both hyperbolic (a) and elliptic (b) cases.
In the presence of an environment, the hyperbolic map loses the recurrence
behavior (solid line) under a Gaussian noise with $\sigma = 0.08$
and maintains a near-constant value, indicating
the ergodicity of the classical map.
On the other hand, the ellptic map still shows recurrence with the same
amount of noise, suggesting classical periodicity.\\
\noindent {\bf Figure 3} $Tr \rho_r^2$ (solid line, left scale) and $<p^2>$
(dashed line, right scale) are plotted
against time for $K = 12$ and $\hbar = 1.52$.
The upper solid line and the lower dashed line correspond to the
case when there is noise, with $\sigma = 0.5$.
As the noise strength increases to $\sigma = 1.5$,
the decoherence time shortens, and
$Tr \rho_r^2$ decays rapidly (the lower solid line).
This accompanies the increase of
diffusive behavior in $<p^2>$ (upper dashed line).\\
\noindent {\bf Figure 4} The von Neumann entropy is plotted versus time
for (a) the quantum kicked rotor with an environment. Here,
$\hbar = 1.52$, $\sigma = 0.08$ and $K = 1.2$.
Entropy stays at zero (reversible dynamics) until a transition regime,
after which the dynamics becomes irreversible.
(b) the quantum cat map, with the same parameters and the same amount of
noise.  We see the same qualitative feature as in the QKR case.

\end{document}